# Impact of surface functionalisation on the quantum coherence of nitrogen vacancy centres in nanodiamond

*R. G. Ryan[1][†], A. Stacey[2] [\*], K. M. O'Donnell[3] T. Ohshima[4], B. C. Johnson[5], L. C. L. Hollenberg[1,2,5,6], P. Mulvaney[1,7], D. A. Simpson[2,6\*]*

[1]School of Chemistry and Bio21 Molecular Science and Biotechnology Institute, The University of Melbourne, VIC 3010, Australia

[†]Now at School of Earth Sciences, University of Melbourne, Parkville, 3010, Australia

[2]School of Physics, University of Melbourne, Parkville, 3010, Australia

[3]Department of Physics, Astronomy and Medical Radiation Science, Curtin University, Bentley, WA, Australia

[4]National Institutes for Quantum and Radiological Science and Technology (QST), Takasaki, Gunma 370-1292, Japan

[5]Centre for Quantum Computation and Communication Technology, University of Melbourne, Parkville, 3010, Australia

[6]Centre for Neural Engineering, University of Melbourne, Parkville, 3010, Australia

[7]Centre for Exciton Science, University of Melbourne, Parkville, 3010, Australia

KEYWORDS: nanodiamonds, nitrogen vacancy, surface functionalisation, quantum properties, spin lattice relaxation, Fourier transform infra-red spectroscopy (FTIR), near-edge X-ray absorption fine structures (NEXAFS).

ABSTRACT

Nanoscale quantum probes such as the nitrogen-vacancy (NV) centre in diamond have demonstrated remarkable sensing capabilities over the past decade as control over the fabrication and manipulation of these systems has evolved. The bio-compatibility and rich surface chemistry of diamond has added to the utility of these probes, but as the size of these nanoscale systems is reduced, the surface chemistry of diamond begins to impact the quantum properties of the NV centre. In this work, we systematically study the effect of the diamond surface chemistry on the quantum coherence of the NV centre in nanodiamonds 50 nm in size. Our results show that a borane reduced diamond surface can on average double the spin relaxation time of individual NV centres in nanodiamonds, when compared to thermally oxidised surfaces. Using a combination of infra-red and x-ray absorption spectroscopy techniques, we correlate the changes in quantum relaxation rates with the conversion of $sp^2$ carbon to C-O and C-H bonds on the diamond surface. These findings implicate double-bonded carbon species as a dominant source of spin noise for near surface NV centres. The link between the surface chemistry and quantum coherence indicates that through tailored engineering of the surface, the quantum properties and magnetic sensitivity of these nanoscale systems may approach that observed in bulk diamond.

INTRODUCTION

In the burgeoning field of quantum sensing, the negatively charged nitrogen-vacancy (NV) centre in diamond[1] has attracted particular interest due to its room temperature quantum coherence and sensitivity to magnetic[2], electric[3], thermal[4] and strain fields[5]. The NV centre in diamond also offers high spatial resolution, limited only by the size of the diamond crystal. Quantum sensing has been demonstrated with NV centres in nanodiamond (ND) in sizes down to 15 nm[6-9], with single NV emission identified in NDs 7 nm in size[10]. However, as the ND size decreases, surface effects begin to hamper the quantum properties of the NV probe. Tetienne *et. al.* have shown that the spin lattice relaxation time ($T_1$) of NV centres in NDs varies by several orders of magnitude over the size range of 20 – 100 nm[9]. This variation is attributed to the spin-spin interactions between the NV centre and surface electronic spins. While this interaction is relatively well described, the origin and source of the surface spin noise is not well understood. This motivates further study of the ND surface to understand the effects of particular moieties on the quantum properties of NV centres.

ND surface chemistry to date, has primarily focused on organic functionalisation methods for reducing ND agglomeration[11, 12] and binding of target molecules such as DNA[13] and biotin[14]. The diamond surface termination has also been explored as a way to modify the fluorescence properties of near surface NV centres[15-17], with the strongest effect observed for hydrogenated diamond. Surface band bending from hydrogenation quenches the fluorescence from NV centres less than 10 nm from the surface of both ND[18] and bulk diamond[19] crystals. This observation highlights the importance of the surface chemistry on not only functionalisation of the nanoparticle, but also the NV luminescence and quantum properties, with a key element being the density of surface spin noise. It has been shown that thermal oxidation provides photo-stable luminescence from NV centres with a surface amenable to functionalisation[20]. The thermally oxidised surface consists of a range of functional groups including carboxyls, hydroxyls, ketones and ethers, $sp^2$ carbon and unsaturated carbon 'dangling bonds'[21-23]. Most functionalisation protocols of NDs proceed via the reduction of surface carbonyl groups using lithium aluminium hydride or borane[14, 24-26] where the hydroxyl groups produced can be further functionalised by silanisation[14, 27, 28] or ester linkages[11, 29]. While these reduction processes are successful from a chemical perspective, their impact on the quantum properties of NV centres has not been studied.

The majority of nanoscale quantum sensing with ND has been achieved using non-detonation ND material sourced from Type Ib high pressure, high temperature (HPHT) bulk material. The surface properties of this material are often considered to be 'bulk like'; however the spin lattice relaxation, $T_1$ time (~40 μs) and transverse dephasing $T_2$ time (~6 μs) from NDs less than 100 nm[7] in size are typically an order of magnitude less than near surface (<25 nm) NV centres in bulk diamond crystals[30], suggesting significant surface-bound sources of magnetic noise. Tisler *et al.* estimated from decoherence measurements ($T_2$) on 30 nm HPHT NDs a density of 13 unpaired spins per square nanometre which is approximately 1/3 of all the surface atoms[31]. Therefore, developing approaches which reduce the density of surface electron spins should give rise to improved ND material.

In this work, we systematically study the effect of surface oxidation and reduction on the quantum properties of single NVs in NDs 50 nm in size. The spin lattice relaxation time, $T_1$, from the same thermally oxidised NDs, is compared before and after borane reduction. Surface analysis of thermally oxidised and borane reduced NDs using a combination of infra-red and x-ray absorption spectroscopies confirms the expected reduction in C=O and C=C content, with a consequential increase in the C-O:C=O ratio and C-$H_x$ species. The borane reduced NDs exhibit a doubling on average in the $T_1$ relaxation time, which is consistent with the observed reduction in double bonded carbon. Our results demonstrate that the quantum and surface chemical properties of these nanoscale systems cannot be treated independently at sizes less than 50 nm and that tailored engineering of the surface can lead to 'bulk like' quantum properties in these nanoscale probes.

RESULTS

**Characterisation of single NV centres in nanodiamond**

The NDs used in this work were synthesised from Type Ib HPHT (Van Moppes) material and exhibited a size range of 50 ± 10 nm when suspended in water from dynamic light scattering (Malvern Zetasizer Nano) . The raw ND powder was electron irradiated and thermally annealed at 1000°C in a $10^{-6}$ Torr vacuum for 2 hours to increase the yield of NV centres per ND (see Methods). The crystal structure of the NV defect is shown schematically in Fig. 1(a). Two electrons are required to form the negatively charged NV centre making it a spin 1 electronic system with a paramagnetic ground state. The ground state energy levels can be read out optically due to the difference in fluorescence intensity of the $m_s = 0$ and $m_s = \pm 1$ sublevels, see Fig. 1(b). The electronic spin of the NV centre is conveniently polarised at room temperature via optical excitation at 532 nm. The $T_1$ relaxation time can be measured by optically polarising the electron spin and allowing it to evolve for a time τ before reading out the spin state with an additional optical pulse, see Fig. 1(c). The resulting $T_1$ decay shown schematically in Fig. 1(d) is dependent on local sources of magnetic noise at frequencies in the GHz range for external magnetic fields less than 100 G. The irradiated and annealed NDs were thermally oxidised in air at 475 °C for 2 hours[32]. The oxidised NDs served as the starting material for the study shown in Fig. 1(e).

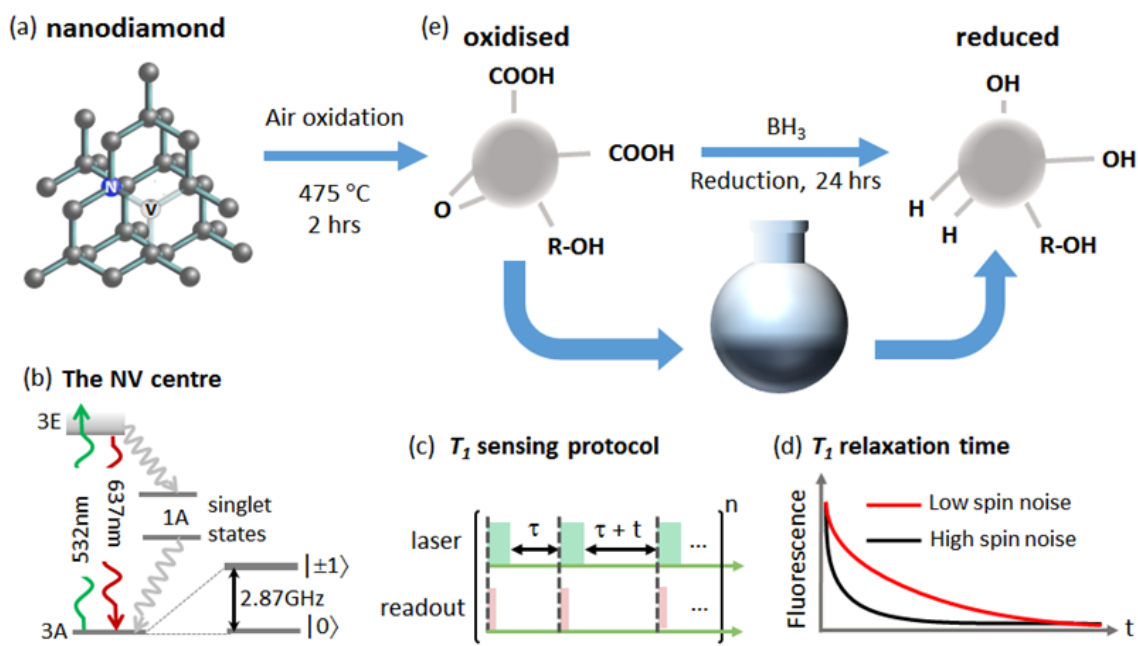

**Figure 1**: Surface modification and quantum spectroscopy of nanodiamond. (a) ND lattice with nitrogen vacancy (NV) defect. (b) A schematic of the energy level structure of the NV centre showing the paramagnetic spin triplet states. (c) Measurement protocol to determine the $T_1$ relaxation time from single NV centres, where τ is the evolution time and n is the number of measurement cycles. (d) A schematic of the resulting $T_1$ relaxation curves with varying levels of magnetic noise. (e) Surface modification procedures: treatment begins with an air oxidation of the ND at 475 °C for 2 hrs, followed by a borane reduction for 24 hours.

For optical imaging, the thermally oxidised NDs were deposited onto a marked coverslip coated with a polyallylamine hydrochloride (PAH) polymer layer to ensure uniformity and to prevent ND agglomeration[6], see Fig. 2(a). Confocal images, shown in Fig. 2(b), were taken of the marked areas with the photo-luminescence (PL) from the NDs analysed using a Hanbury-Brown and Twiss interferometer to identify NDs hosting single NV centres. The PL from each ND was then monitored over a 5 min period to ensure photo-stability. The $T_1$ relaxation times of ten photo-stable NV centres were measured sequentially, in an 80 × 80 μm field of view, see Fig. 2(c). After characterisation of the thermally oxidised NDs, the NDs underwent a borane reduction by placing the coverslip containing the NDs into a solution of borane tetrahydrofuran (see Methods). The characterisation of the same ten individual NDs was then repeated.

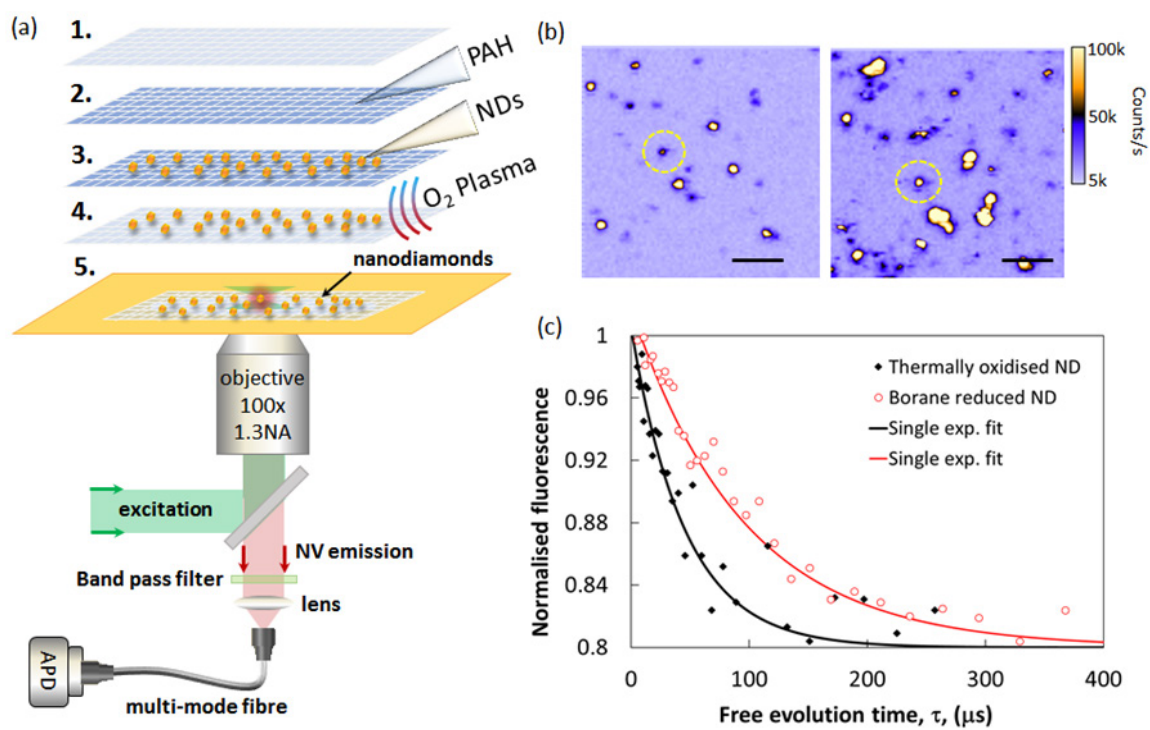

**Figure 2**: (a) Schematic diagram of the deposition process to prepare ND samples for confocal imaging. Once the NDs were on the glass coverslip, the borane reduction was re-run with the whole slide in the reaction solution. (b) Confocal images showing a particular set of NV centres before and after borane reduction, indicating the ease of re-identification post-reaction; the black scale bar represents 5 μm. (c) Spin relaxation ($T_1$) curves for a thermally oxidised and borane reduced ND (NV #8) obtained from n = $10^6$ measurement cycles.

The $T_1$ relaxation times for the oxidised and reduced NDs are shown in Fig. 3(a) with the percentage differences between thermally oxidised and borane reduced NDs shown in Fig. 3(b).

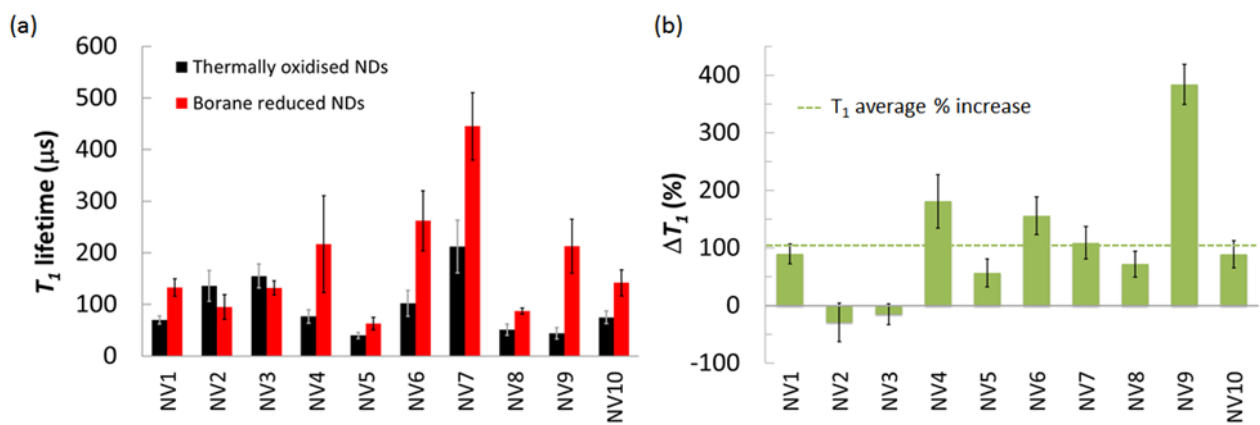

**Figure 3**: Spin lattice relaxation times for thermally oxidised and borane reduced 50 nm NDs. (a) Direct $T_1$ time comparison between the same set of NDs which were thermally oxidised and then borane reduced. (b) Percentage change in $T_1$ after the borane reduction. Error bars in both plots are determined by the error in the single exponential fit to the $T_1$ decay curves. The data shows an average increase in the $T_1$ time of 110% over the ten individual NDs studied.

The NV centres exhibited on average a doubling of the $T_1$ time across the ten individual NDs. The variability of the increases can be attributed to the fact that the reaction was performed on NDs adhered to the coverslip. Therefore, depending on the exact position of the NV centre within the ND, it may not experience a significant change in local surface state. For example, a NV centre residing at the bottom of the ND close to the coverslip will be dominated by surface spin noise at the diamond/coverslip interface which in this case is protected from the chemical reduction. However, given that 80% of the NDs experienced a statistically significant increase in $T_1$ time, the surface modification clearly has a favourable impact on the spin noise spectrum. To understand the specific role that borane reduction has on the quantum properties of the probes we employed infra-red and X-ray absorption spectroscopy techniques.

**Spectroscopic characterisation of nanodiamonds**

The Fourier-transform infrared spectroscopy (FTIR) spectrum of the oxidised surface prior to reaction with borane is shown as a black trace in Fig. 4(a). There is a broad peak centred at around 3400 cm$^{-1}$ in Fig. 4(a) which is characteristic of the O-H stretch. This is attributed to a combination of adsorbed water on the surface of the ND, carboxylic O-H groups or bonded hydroxyls. This is matched by a peak at 1630 cm$^{-1}$, which we assign to the C=C stretch[33]. The largest features are a C=O peak centred at 1780 cm$^{-1}$ which encompasses the frequencies of carbonyl-containing groups such as ketone, esters, carboxylic acids and carboxylic anhydrides and a C-O peak at 1100 cm$^{-1}$. Following the collection of the thermally oxidised spectrum, borane reduction was performed as described in the Methods. The borane reduction brings out several new peaks in the fingerprint region (1500-500cm$^{-1}$), see the red trace in Fig. 4(a), which we assign to oxygen containing groups such as ethers and alcohols[34]. The peak at 1630 cm$^{-1}$ is removed after the borane reduction which is attributed to the conversion of C=C to sp$^3$ groups and evidenced by the appearance of a C-H$_x$ peak at 2962 cm$^{-1}$. However, many reports have discussed the problematic nature of assigning hydroxyl FTIR peaks from NDs, due to the ease of water adsorption at the surface[14, 26, 32, 35]. In search of more surface-specific information on the borane reduction process, we utilized near-edge X-ray absorption fine structures (NEXAFS) spectroscopy.

Figure 4(b) shows the oxygen K-edge NEXAFS spectra of the oxidised (in black) and borane reduced (in red) NDs, after annealing *in-vacuo* to remove any water contamination. The features of note are the pre-edge peak at 532 eV, assigned to sp$^2$ oxygen in C=O, and at 540 eV, assigned to C-O[32]. The C=O peak is greatly reduced following borane reduction, in agreement with the FTIR spectra. The consequential increase in the C-O:C=O ratio is a good indicator of the efficacy of the borane reduction process in reducing carbonyl groups, and likely producing hydroxyl groups.

Figure 4(c) presents the carbon K-edge NEXAFS spectrum for the thermally oxidised, and borane reduced ND samples. The key features of this extended spectrum are the diamond core-hole excitonic peak at 289 eV and the second band gap of diamond at approximately 302 eV[32]. In the case of oxidation and reduction, this second band gap dip is preserved, indicating that these reactions are not adding extra (non-diamond) carbon to the surface. A zoomed-in section showing the pre-edge features of the carbon K-edge NEXAFS spectrum is shown in Fig. 4(d). This pre-edge detail allows surface-specific analysis of many of the functional groups present. Peaks at 285 eV are generally assigned to sp$^2$ (C=C) groups, and peaks around 286.5 to carbonyls (C=O) [36]. The NEXAFS peak at 287.4 eV is generally assigned to C-H[36, 37]. Krueger and Lang have stated that borane reduction followed by an acidic work-up is likely to produce C-H bonds from $sp^2$ carbon moieties, supporting the finding of decreased C=C and increased C-H[24]. A very small peak is evident in the C-NEXAFS spectrum for the oxidised sample at ~282 eV, which can be assigned to dangling bonds[38]. However, we ascribe this to the *in-vacuo* annealing induced loss of weakly bonded moieties on the surface, revealing dangling bonds, rather than an indicator of dangling bonds under ambient conditions.

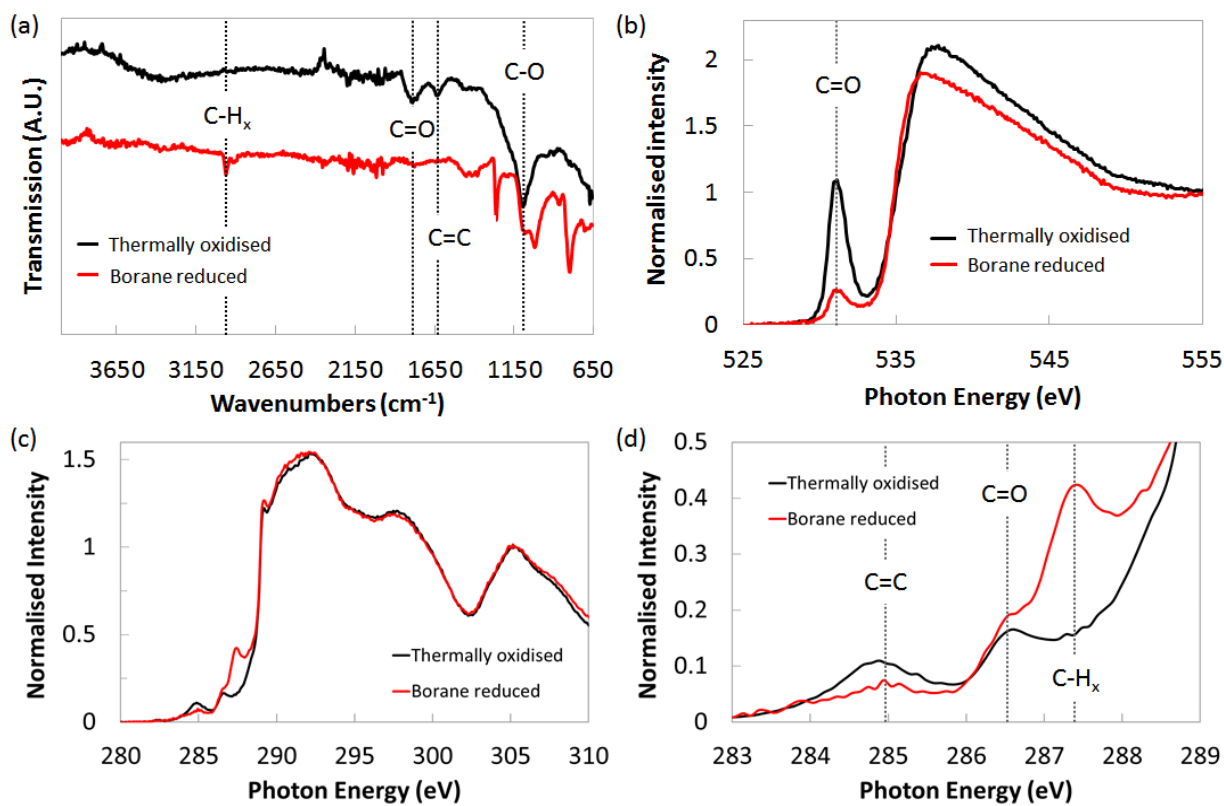

**Figure 4**: (a) FTIR spectra for the oxidised, borane reduced NDs, taken on an ATR plate FTIR spectrometer. (b) Oxygen K-edge NEXAFS spectra, taken on the soft x-ray beamline at the Australian Synchrotron, of the oxidised and borane reduced NDs. This demonstrates a clear reduction in carbonyl C=O bonds as a result of the borane reduction. (c) Carbon K-edge NEXAFS spectra, also taken at the Australian Synchrotron, of oxidised and reduced NDs. (d) Zoomed in section of the carbon NEXAFS spectrum, showing pre-edge functional detail.

Overall, the key findings from the spectroscopic characterisation are (1) borane reduction causes a loss of C=O and an associated production of C-O groups, as evident from both oxygen NEXAFS and FTIR and (2) borane reduction produces C-H groups by reacting with $sp^2$ carbon, including C=C, as seen by carbon NEXAFS and FTIR. These findings allow insight into the mechanisms responsible for the increase in spin lattice relaxation time for the borane reduced NDs. A hypothesis to explain the $T_1$ increase is that replacing C=O and C=C groups with C-O or C-H groups reduces the density of surface electron traps with unoccupied states low in the diamond band-gap. This modifies the spectral characteristics of the surface spin noise and also changes the pathways and probabilities of NV ionization processes, leading to an increase in the spin-relaxation time.

CONCLUSION

The utility of NDs for quantum sensing applications is determined by both their physical and chemical characteristics. In this work, we demonstrate that these factors must not be considered independently. We show spectroscopically that the surface chemistry of NDs can be modified to reduce the amount of $sp^2$ carbon, and subsequently reduce the surface spin noise. Following a borane reduction, which is often the first step in functionalisation, we observe on average a doubling of the spin lattice relaxation time for 80% of the individual NDs studied. The $T_1$ time of NV centres is an important resource for quantum sensing[39, 40], by increasing the coherence times in these systems we can improve the magnetic sensitivity for nanoscale magnetometry applications. With further improvements in the surface modification, sub-10 nm NDs with 'bulk like' coherence properties may be within reach, which would directly enhance the array of sensing applications of NDs already underway. Further screening of functionalisation pathways and target moieties will also be required to understand the effect that each group has on NV quantum properties, prior to their use as targeted functionalised quantum probes.

METHODS

**Nanodiamond preparation**

NDs were purchased from VanMoppes (SYP 0–0.01) and irradiated with high-energy electrons (2 MeV with a fluence of $1 \times 10^{18}$ electrons/cm$^2$) and vacuum annealed at 1000 °C for 2 hrs at the National Institutes for Quantum and Radiological Science and Technology (Takasaki, Japan). Thermal oxidation was performed by placing 10 mg of NDs into a ceramic asher which was heated to 475 °C in air for 2 hrs. Oxidised and reduced NDs were dispersed in water at a concentration of 1 mg/ml and sonicated for 1 hour. The suspension was centrifuged at 12,000 g for 120 s and the supernatant was removed and used as stock suspension. For adsorption onto NDs, cleaned marked glass substrates (No. 1, Menzel) were immersed in a 1 mg/mL solution of polyallylamine hydrochloride (PAH) (70 kDa) in 0.5 M NaCl in Milli-Q water for 5 min. After rinsing with Milli-Q water, the stock ND suspension was diluted × 50 in Milli-Q water and then applied to the coverslip for 5 min, before rinsing again. The glass substrates were treated in a 40-W oxygen plasma (25% oxygen in argon, 40 standard cubic centimetres per minute (sccm) flow rate) for 5 min to remove the polyelectrolyte from the substrate for the oxidised samples only.

**Spectroscopic characterisation**

Fourier transform infrared (FTIR) spectra of ND powders were recorded on an attenuated total reflection (ATR) spectrometer. Approximately 1 mg of powder was required for each measurement. For near-edge x-ray absorption fine structure (NEXAFS) measurements, powder samples were placed in water, and evaporated on a gold/platinum/titanium/glass substrate. Spectra were recorded in the ultra-high vacuum endstation of the soft X-ray beamline at the Australian Synchrotron. NEXAFS measurements were conducted in an analysis chamber with a base pressure of $< 1 \times 10^{-10}$ mbar. NEXAFS measurements were measured in the 270–320 eV and 520–555 eV ranges, for carbon K-edge and oxygen K-edge regions respectively. The NEXAFS signal was recorded using a channeltron with a retarding field grid, set to 230 eV for the carbon NEXAFS scans. This setup allows for a partial electron yield (PEY) NEXAFS signal, based on Auger electrons which originate within ~1 nm from the surface, thus allowing accurate determination of the near-surface electronic environment of each sample. Measurements were recorded before and after annealing at ~370 °C, in order to desorb any water and adventitious carbon. All measurements utilised a flood gun for charge stabilisation under X-ray illumination, however this was found to be insufficient for neutralising the oxidised ND sample surface prior to annealing *in-vacuo*. As such the carbon K-edge NEXAFS data presented for the oxidised sample is for its annealed state, whereas the pre-anneal data are presented for the borane reduced samples.

**Borane reduction**

40 mg of air oxidised ND powder for spectroscopic characterisation was placed in a round bottomed flask with a magnetic stirrer bar and condenser. This was securely stoppered, placed on a standard Schlenk line and purged with $N_2$ gas. 5 mL 1 M $BH_3$.THF solution was added by syringe and the mixture stirred and refluxed at 67 °C for 24 hours. The reaction was quenched by the addition of 2 M HCl until $H_2$ gas evolution ceased. The NDs were separated from the reaction mixture by centrifugation. The powder was rinsed in acetone, then water, followed by stirring in water for two hours. The ND powder formed a stable suspension in water, but was removed by centrifugation for analysis or further reactions. The same borane reduction procedure was followed, without centrifugation, for NDs attached to the cover slip for confocal analysis.

**Confocal imaging**

Confocal microscopy measurements were performed on a custom built inverted microscope (Nikon Ti-U), with a 100x 1.3NA oil-immersion objective (Nikon). Optical excitation was provided by a laser at 532 nm (Verdi, Coherent Scientific). The power density of the optical excitation was kept below 100 µW/µm$^2$ to prevent photo-ionisation of the single NV centres. The laser was modulated via an acousto-optic modulator (Crystal technologies) and controlled by a PulseBlaster card (Spincore). The NV fluorescence was filtered using a long pass filter at 560 nm in combination with a bandpass filter from 650-750 nm (Semrock). The filtered fluorescence was detected using a single photon counting avalanche photodiode (Perkin Elmer), with the time correlated photons captured with a Multiple-Event

Time Digitizer card (Fast ComTech P7889). The microscope and data acquisition were controlled via custom LabVIEW code.

AUTHOR INFORMATION

**Corresponding Author**

*simd@unimelb.edu.au and alastair.stacey@unimelb.edu.au.

**Author Contributions**

D.A.S., A.S. and L.C.L.H. conceived the study. D.A.S. designed and constructed the quantum sensing microscope and supervised the project. T. O. and B. C. J. performed the electron irradiation and high temperature annealing of the ND material used in the work. R. R. deposited the NDs samples for imaging, acquired the $T_1$ data, performed the thermal oxidation and borane reduction with assistance from D. A. S. and P.M. The FTIR spectra were acquired and analysed by R. R. and A. S. and K. M. O performed the x-ray absorption spectroscopy measurements. A. S. analysed the x-ray absorption data. All authors discussed the results and participated in writing the manuscript.

## ACKNOWLEDGEMENT

The authors acknowledge fruitful discussions with Dr Jean-Philippe Tetienne and Dr Liam Hall. Spectroscopic analysis for this research was undertaken on the Soft X-ray Spectroscopy beamline at the Australian Synchrotron, Victoria, Australia. This research was supported in part by the Australian Research Council Centre of Excellence for Quantum Computation and Communication Technology (Project number CE110001027). This work was also supported by the University of Melbourne through the Centre for Neural Engineering and the Centre for Neuroscience. L.C.L.H acknowledges support of the Australian Research Council under the Laureate Fellowship scheme (FL130100119).